# Printing out Particle Detectors with 3D-Printers – a Potentially Transformational Advance for HEP Instrumentation

M. Hohlmann

*Dept. of Physics and Space Sciences, Florida Institute of Technology, Melbourne, FL 32901*

September 3, 2013

*Abstract* – **This white paper suggests posing a "grand challenge" to the HEP instrumentation community, i.e. the aggressive development of additive manufacturing, also known as 3D-printing, for the production of particle detectors in collaboration with industry. This notion is an outcome of discussions within the instrumentation frontier group during the 2013 APS-DPF Snowmass summer study conducted by the U.S. HEP community. Improvements of current industrial 3D-printing capabilities by one to two orders of magnitude in terms of printing resolution, speed, and object size together with developing the ability to print composite materials could enable the production of any desired 3D detector structure directly from a digital model. Current industrial 3D-printing capabilities are briefly reviewed and contrasted with capabilities desired for printing detectors for particle physics, with micro-pattern gaseous detectors used as a first example. A significant impact on industrial technology could be expected if HEP were to partner with industry in taking on such a challenge.**

## I. Introduction

Additive manufacturing or "3D-printing" was developed by industry over the past decades for rapid prototyping or small-scale production of machine parts during the development phase of engineering projects. 3D-printers can build up three-dimensional objects of any shape, most commonly by extruding small beads of thermoplastic material or a photopolymer from a nozzle or syringe and precisely placing the material in layers as specified by a digital model. The object is then built up layer by layer. The iconic open-source RepRap[1] printer, a self-replicating printer capable of printing its own components, and the commercial "Replicator" printer series from MakerBot[2] were embraced early on by technology enthusiasts. The technology has now reached the consumer market with low-cost desktop 3D-printers available from a number of vendors at prices below $2,000, such as the "Cube" from 3D Systems[3], or the "Creatr" from Leapfrog[4].

The complementary technique of direct metal laser sintering also allows 3D-printing of solid metal objects. In this process, a laser melts layers in a bed of granular metal powder fusing them into a solid metal object of desired shape. Selective laser sintering of powder mixtures or coated powder allows additive manufacturing of objects from a variety of commercially available powder materials, such as polymers, metals, and alloys.

## II. Current 3-D printing capabilities

**Materials:** Popular thermoplastic polymers extruded by 3D-printers are acrylonitrile butadiene styrene (ABS), the material that "Lego" bricks are made of, high and low density polyethylene (HDPE, LDPE), polypropylene (PP), unplasticized polyvinylchloride (UPVC), and polylactic acid (PLA). A variety of metals, such as stainless steel, aluminum, or titanium can be printed with the direct metal laser sintering process. 3D-printers for biotech application, such as the "3D-Bioplotter" from EnvisionTEC[5], can build up scaffolds for growing tissue by printing a variety of biomaterials, such as hydrogels, polymers, ceramics, and metals, while direct 3D printing of biological cells is also under development[6] with the ultimate goal of printing human replacement organs for medicine.

**Resolution and accuracy:** Currently available high-definition commercial 3D printers, such as the ProJet HD 7000 from 3D Systems[3], achieve a resolution of 75 microns in the x-y print plane while the thickness of each print layer in the z direction is 50 microns. Specialized printers for biomaterials [5] can achieve one micron resolution albeit over a limited build space. Typical placement accuracies are 0.1%.

**Build volumes and print speed:** The larger build volumes of commercial printers currently reach dimensions x × y × z ≈ 50 cm × 50 cm × 25 cm. Print speeds are on the order of 10 cm/s and it typically takes minutes to hours to print out an object depending on size and complexity.

## III. R&D challenges for 3D-printing micro-pattern particle detectors

Since the author currently works on HEP applications of micro-pattern gas detectors (MPGD's)[7], specifically gas electron multipliers (GEMs)[8], this detector class is used for demonstrating what advances would be required of 3D-printing technology to apply it to MPGD's and what could be gained if it were possible to print MPGD's. Similar reasoning applies to other types of detectors that use microscopic patterns, e.g. micro-channel plates.

The diameter of a hole in a standard GEM is 70 μm and the hole pitch is 140 μm, so the resolution of a 3D printer would have to be considerably smaller, i.e. on the order of 1 μm, to cleanly build up a 70 μm structure. Similarly, a GEM foil is 60 μm thick, so the thickness of the build layers would also need to be reduced to about 1 μm to build up a proper vertical cross section for the hole structure. With blob size decreasing, print speeds would have to increase correspondingly, about two orders of magnitude, to keep a print-out process reasonably short.

A crucial requirement would be the ability to directly print composite materials. A GEM foil is made from copper-clad kapton, so a 3D-printer would have to be able to print a polymer on top of a metal and vice versa to create the desired structure. To achieve this, the currently separate 3D-printing technologies for polymers and metals would have to be integrated. A first

**Table 1.** *Typical current commercial 3D-printing capabilities contrasted with performance goals desired for 3D-printing micro-pattern gas detectors for HEP.*

|  | **Current capability** | **Performance goal** |
|---|---|---|
| **Printing resolution in x-y** | ~ 75 μm | ~ 1 μm |
| **Layer thickness in z** | ~ 50 μm | ~ 1 μm |
| **Print speed** | 10 cm/s | > 100 cm/s |
| **Materials** | Either polymers or metals | Polymer-metal composites |
| **Object size** | 50 cm × 50 cm × 25 cm | 200 cm × 100 cm × 10 cm |

step in this direction might be the use of conductive polymers instead of metals where possible.

Many detector systems in HEP are constructed from layers of detection elements, some of which need to cover rather large areas, e.g. in a muon system. Consequently, the base area for 3D printers would have to be increased to cover more than a square meter; at the same time the build-up in the vertical direction could be reduced somewhat for a flat large-area detector. The various performance goals are contrasted with existing 3D-printing capabilities in Table 1.

**IV. Expected gains for detector technology from advanced 3D-printing capabilities**

If all the above problems were solved, what could be gained for particle detectors from 3D-printing technology? Once any 3D structure down to the micron level could be created almost instantaneously for any combination of materials, then any particle detector structure that a creative detector designer can imagine could simply be "printed out" almost immediately. All that would be required is for the structure to be designed as a digital 3D-model with a CAD program. Staying with the MPGD example, many innovative gas amplification structures have been devised for these detectors in the past and with advanced 3D-printing improved structures with even better performances will presumably emerge. From a long-term perspective, a tool this powerful could potentially transform the design, development, and production of future particle detectors. The possibilities for innovative particle detector designs might indeed be endless.

Instead of producing individual parts for detectors that then get assembled into the complete detector, with this technology an entire detector could be printed out in one go as a self-contained unit. For example, to build a Triple-GEM detector, currently individual GEM foils are produced in a complex (and expensive) chemical etching process of copper-clad kapton base material. Because they are not rigid bodies, the foils must be stretched thermally or mechanically by hand and are then mounted inside a gastight chamber. This process makes the detectors relatively expensive. Instead, with advanced 3D-printing technology, it might be possible to simply print out an entire detector with GEM-like structures inside a chamber as a single unit.

While deep sub-micron structures as used in integrated circuits are probably too small to be printed directly, discrete electronic components would be large enough to be printed out in an advanced 3D-printing process. This would open the door to printing detector electronics, e.g. for readout, control, or high voltage, directly onto or even into a detector.

From a manufacturing point of view, advanced 3D-printing technology could put the manufacturing tools directly into the hand of the end user, i.e. the detector physicist. Instead of relying on a small number of institutes or workshops with specialized know-how that can handle the required manufacturing technology, any user with access to an advanced 3D-printer, even if only via a remote advanced 3D-printing service, could become a small-volume producer of detectors. This should speed up the turn-around in the development phase. It should also bring down the cost for detector prototyping since the unit price for 3D-printing a small number of detectors should not be too different from the price for a mass-produced printed unit.

Mass production itself could be affected positively with the ability to submit 3D-print jobs to many printers – local or remote – simultaneously. This could get the community around typical problems with industrialization of emerging detection technology, where industry is often not very keen on making investments when the initial market is confined to HEP and consequently rather small and not very profitable. For example, using again the case of MPGD's, the MPGD community has been struggling for years to transfer MPGD manufacturing technology from CERN to industry. Even when progress is made with a company, experience

SNOW13-00137shows that it is typically a slow process that also can end abruptly. In the end, manufacturing via 3D-printing should be able to bring down the overall cost of mass-manufacturing detectors for particle physics experiments.

One could expect advanced 3D-printing technology to have a serious technological impact beyond HEP. To give an example that is a straightforward extrapolation from the above: If any 3D structure down to the micron level could be created easily for any combination of materials, then "3D-pcb's" featuring directly integrated electronics components could be printed out by anyone who can design them, not just by specialized printed circuit board factories.

## V. Conclusion

Given that 3D-printing has the potential for transforming HEP detector technology, it is suggested that the HEP community consider making an aggressive effort to develop advanced 3D-printing in cooperation with industry by investing appropriate financial and intellectual resources. This effort could be facilitated if the HEP funding agencies were to pose the development of 3D-printed detectors as a "grand challenge" to the instrumentation community, e.g. by funding cooperation between 3D-printing industry and HEP via SBIR programs.

## VI. Acknowledgments

This paper is inspired by the author's informal discussions on 3D-printing with C. Haber (LBNL), P. Kim (DOE), H. Nicholson (Mt. Holyoke), P. Wilson (FNAL), and others during the 2013 Snowmass community summer study at the U. of Minnesota in Minneapolis. The author would like to thank these individuals for their input and in particular H. Nicholson for his encouragement to write this up in this white paper. He is also grateful to the "Snowmass on the Mississippi" organizers at the U. of Minnesota for hosting a well-organized productive meeting.

## References

[1] Information available online at http://reprap.org/.

[2] Information available online at http://www.makerbot.com/.

[3] Information available online at http://www.3dsystems.com/.

[4] Information available online at https://www.lpfrg.com/product/creatr/.

[5] Information available online at http://envisiontec.com/.

[6] B. Derby, "Printing and prototyping of tissues and scaffolds," Science, 338 (2012) 921-926.

[7] M. Titov, et al., "Micro-pattern gaseous detector technologies and RD51 collaboration," Modern Physics Letters A, 28 (2013).

[8] F. Sauli, "GEM: A new concept for electron amplification in gas detectors," Nucl. Instr. and Meth. A, 386 (1997) 531-534.4